\documentclass[sigconf]{acmart} 

\usepackage{booktabs, enumitem}

\AtBeginDocument{%
  \providecommand\BibTeX{{%
    \normalfont B\kern-0.5em{\scshape i\kern-0.25em b}\kern-0.8em\TeX}}}

\copyrightyear{2021} 
\acmYear{2021} 
\setcopyright{acmlicensed}\acmConference[MMAsia '21]{ACM Multimedia Asia}{December 1--3, 2021}{Gold Coast, Australia}
\acmBooktitle{ACM Multimedia Asia (MMAsia '21), December 1--3, 2021, Gold Coast, Australia}
\acmPrice{15.00}
\acmDOI{10.1145/3469877.3490612}
\acmISBN{978-1-4503-8607-4/21/12}

\begin{document}

\title{Score Transformer: Generating Musical Score\\ from Note-level Representation}
\renewcommand{\shorttitle}{Score Transformer: Generating Musical Score from Note-level Representation}


\author{Masahiro Suzuki}
\affiliation{%
  \institution{Yamaha Corporation}
  \city{Shizuoka}
  \country{Japan}}
\email{masahiro1.suzuki@music.yamaha.com}

\renewcommand{\shortauthors}{Masahiro Suzuki}

\begin{abstract}
In this paper, we explore the tokenized representation of musical scores using the Transformer model to automatically generate musical scores. Thus far, sequence models have yielded fruitful results with note-level (MIDI-equivalent) symbolic representations of music. Although the note-level representations can comprise sufficient information to reproduce music aurally, they cannot contain adequate information to represent music visually in terms of notation. Musical scores contain various musical symbols (e.g., clef, key signature, and notes) and attributes (e.g., stem direction, beam, and tie) that enable us to visually comprehend musical content. However, automated estimation of these elements has yet to be comprehensively addressed. In this paper, we first design score token representation corresponding to the various musical elements. We then train the Transformer model to transcribe note-level representation into appropriate music notation. Evaluations of popular piano scores show that the proposed method significantly outperforms existing methods on all 12 musical aspects that were investigated. We also explore an effective notation-level token representation to work with the model and determine that our proposed representation produces the steadiest results.
\end{abstract}

\ccsdesc[500]{Applied computing~Sound and music computing}

\keywords{symbolic music representation, music transcription, musical score generation, transformers}


\maketitle

\section{Introduction}

\begin{figure}[t]
 \centerline{
 \includegraphics[width=1.0\columnwidth]{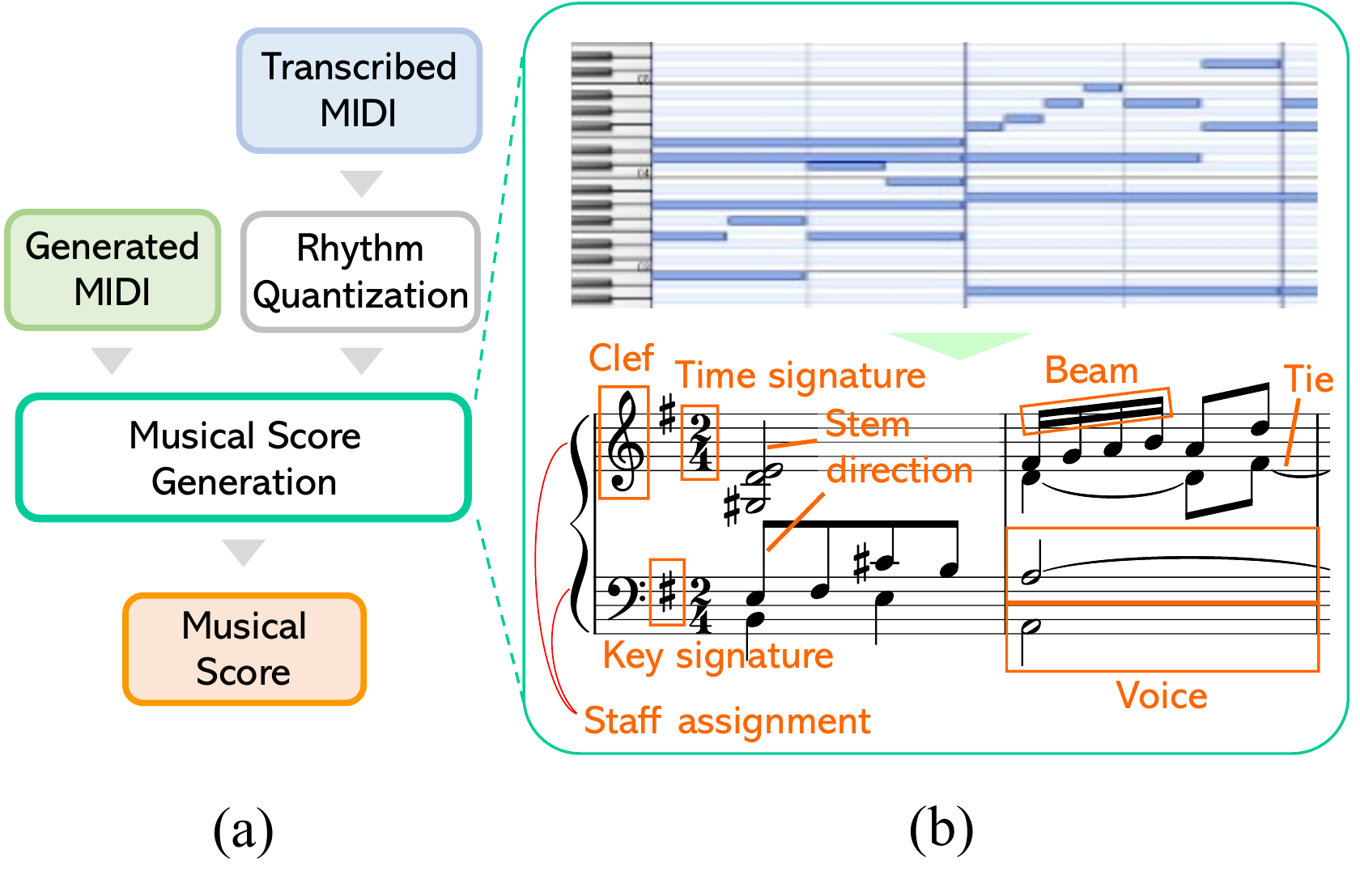}}
 \caption{Overview of (a) MIDI-to-score process and (b) musical score generation task. We focus on (b), which generates human-readable music notation (lower) out of quantized note-level representation (upper).}
 \label{fig:overview}
\end{figure}

Musical scores are symbolic visual media that contain rich musical information necessary to read and play music on musical instruments. Various musical symbols (e.g., time and key signature) and attributes (e.g., voice and grouping of notes) in musical scores enable us to comprehend music easily. In contrast, note-level representations (e.g., piano-rolls and MIDI files) cannot express such high-level visual musical information: they only represent music aurally. Considering the differences, it is of great importance to develop automated systems able to generate representations equivalent to musical scores.

Deep learning techniques have yielded impressive results in music generation and music transcription. However, their application to the generation of comprehensive musical scores or even effective representations thereof remains unexplored. In music generation, sequence models with symbolic music representation have exhibited considerable success \cite{Huang2020, Oore2018}. In particular, the combination of the Transformer model \cite{Vaswani2017} and a token representation of music has been revealed to generate coherent and structural music \cite{Huang2019, Ji2020}. However, the music representations used in prior studies were note-level (MIDI-equivalent); hence, their expressions were limited compared to those of musical scores. With regard to automatic music transcription (AMT), neural network models have achieved remarkable success in the audio-to-MIDI process (e.g., multi-pitch estimation and onset/offset detection) \cite{Benetos2019}. In contrast, the subsequent MIDI-to-score process in which note-level representation is transcribed into music notation \cite{Foscarin2019}, has not been comprehensively addressed \cite{Cogliati2016}. More specifically, the MIDI-to-score process comprises two subtasks \cite{Foscarin2019, Nakamura2018, Shibata2020} as illustrated in Figure \ref{fig:overview}(a). There has been a lack of focus on the latter musical score generation (or score typesetting) subtask, in which rhythm-quantized (beat-estimated and quantized) MIDI is transcribed into a comprehensive musical score. Although notation-level symbolic representations have been partially employed in a few studies \cite{Carvalho2017, Liu2021, Roman2018, Roman2019}, using those of comprehensive musical scores comprising various musical symbols and attributes with sequence models has yet to be explored. Therefore, the issue of whether sequence models can generate comprehensive music notations and of how to effectively represent musical scores to feed such models remain unresolved.

In this study, we address the generation of comprehensive musical scores using the Transformer model \cite{Vaswani2017}, which complements music generation and music transcription research. This task is equivalent to the musical score generation subtask in music transcription and also aids music generation by transcribing its note-level output into musical scores (Figure \ref{fig:overview}(a)). We first design token representations corresponding to musical scores (Section \ref{sec:3}). Then, we train and evaluate Score Transformer, a Transformer model that translates quantized note-level representations into appropriate music notation (Sections \ref{sec:4}–\ref{sec:6}).
Figure \ref{fig:overview}(b) illustrates a variety of subtasks required to notate music, such as staff assignments of notes, key estimation, and voice segregation of the note stream. We address this challenging and important task while addressing the research issues presented above.

Our contributions are summarized as follows:
\begin{itemize}[leftmargin=1.25em]
\item We propose a novel method to generate comprehensive musical scores, which jointly estimate various musical symbols and attributes required in musical scores.
\item We propose an effective token representation of music notation that works well with the Transformer model. To promote further research, tools designed to convert between the proposed representation and MusicXML \cite{Good2003}, a standard open score exchanging format, are provided.\footnote{\url{https://github.com/suzuqn/ScoreTransformer/}}
\item We devise a simple training scheme for bridging two levels of symbolic musical representation: \textit{note}-level and \textit{notation}-level. This scheme requires only musical score data.
\end{itemize}

\section{Related Work}
\subsection{Musical Score Generation}
Musical information should be encoded in a form that an employed model can interpret. For this purpose, the tokenization of musical information is a common choice for sequence models. For note-level (MIDI) tokenization, two representation styles have been proposed: MIDI-like \cite{Oore2018} and REMI \cite{Huang2020}. The former style represents notes with \textit{note-on, note-off}, and \textit{delta-time}, which are similar to MIDI events, whereas the latter style includes \textit{bar, position}, and \textit{note duration} tokens to improve the former representation. In REMI, a set of \textit{note-on, note-off} events of the same note can be represented as a single token, which resembles note value in musical scores. Recent research has also revealed that the sequence model can generate note-level representations in an efficient form where related tokens are grouped \cite{Hsiao2021}. In this paper, we design notation-level token representations that correspond to a musical score. We also expand REMI to various meters for our note-level representation.

\subsection{Note-level Tokenization}
In a few prior studies on music transcription, the generation of musical scores has been attempted in a limited manner. Two types of studies have been conducted: MIDI-to-score and audio-to-score (in an end-to-end manner). The former studies \cite{Cogliati2016, Foscarin2019} addressed the MIDI-to-score task with a probabilistic model or grammar model. For example, a framework was proposed in \cite{Cogliati2016} that transcribes performance MIDI into a musical score using LilyPond \cite{Nienhuys2003} by splitting the task and solving them using a probabilistic analysis tool \cite{Temperley2009}. The latter studies \cite{Carvalho2017, Liu2021, Roman2018, Roman2019} attempted to output music notations directly from waveforms. For example, a recurrent neural network (RNN) or its variant has been utilized to generate LilyPond \cite{Carvalho2017, Liu2021}, Humdrum \cite{Roman2019}, and designed characters \cite{Roman2018}. These studies have shown the possibility of automated generation of musical scores with RNNs; however, they considered only a limited set of musical symbols. In this paper, we aim to generate comprehensive musical scores in the form of token representation with the Transformer model and explore their suitable representation. We focus on piano scores, the typical musical scores with multiple staves in a part, and multiple voices in a staff. 

\section{Score Tokenization}
\label{sec:3}
We address score tokenization in two ways: 1) by designing new token representations, and 2) by using existing score formats.

\subsection{Proposed Representation}
Inspired by the success of note-level tokenization, we design a token representation that symbolizes score elements. Our design principles are as follows:

\begin{itemize}[leftmargin=1.0em]
\item{\textbf{One token per symbol or attribute:}} A single token corresponds to a score symbol (e.g., barline, clef, key signature, and time signature) or a note attribute (e.g., note value, stem direction, and tie). The only exception is a voice symbol, which consists of a pair of tokens (Section \ref{sec:voice}). Because following this principle results in long sequences, we also consider a shorter form (Section \ref{sec:concat}).
\item{\textbf{Compatible with music21 attributes:}} The attributes of the symbols are designed to be compatible with those of music21 \cite{Cuthbert2010} objects, thus enabling utilization of the music21 toolkit to create MusicXML scores from the token representation.
\item{\textbf{Concatenated sequences of staves:}} We concatenate time-ordered token sequences of staves (i.e., right- and left-hand parts of piano scores) to form a single sequence \cite{Ens2020}, which allows the model to refer to generated tokens while making inferences. This helps the model maintain the consistency of score attributes and staff assignments between staves.
\item{\textbf{Tokenize essential musical symbols only:}} We represent the essential musical symbols in musical scores listed in Table \ref{table:symbols}. Additional expression symbols (e.g., articulations, dynamics, and ornaments) and repeat symbols are not included in this work.
\end{itemize}

\begin{table}[h]
 \caption{Symbols and their variations in the proposed score token representation.}
 \label{table:symbols}
  \centering
  \small
  \renewcommand{\arraystretch}{1.35}
    \begin{tabular}{lcc}
    \toprule
    \textbf{Symbol} & \textbf{Example} & \textbf{Variations} \\
    \midrule
    Staff & \textit{R}     & \textit{R/L} \\
    Barline & \textit{bar}   & \textit{bar} \\
    Clef  & \textit{clef\_treble} & \textit{clef\_\{bass/treble\}} \\
    Key Signature & \textit{key\_flat\_2} & 
    \begin{tabular}{c} \textit{key\_\{sharp/flat/natural\}} \\ \textit{\_\{1, 2, …, 6\}} \end{tabular} \\
    Time Signature & \textit{time\_4/4} & \textit{time\_\{2/4, 3/4, 4/4, etc.\}} \\
    Voice & \textit{<voice>} & \textit{<voice>, </voice>} \\
    Rest  & \textit{rest}  & \textit{rest} \\
    Pitch & \textit{note\_C4} & 
    \begin{tabular}{c} \textit{note\_\{A, B, ..., G\}} \\ \textit{\{\#\#/\#/b/bb/(none)\}\newline{}\{0, 1, …, 8\}} \end{tabular} \\
    Duration & \textit{len\_1/2} & \textit{len\_\{1/24, 1/16, …, 4\}} \\
    Stem Direction & \textit{stem\_up} & \textit{stem\_\{up/down\}} \\
    Beams & \textit{beam\_stop} & 
    \begin{tabular}{c} \textit{beam\_\{start/stop/continue/} \\ \ \ \ \textit{partial-left/partial-right\}\_...} \end{tabular} \\
    Tie   & \textit{tie\_start} & \textit{tie\_\{start/continue/stop\}} \\
    \bottomrule
    \end{tabular}%
    \renewcommand{\arraystretch}{1}
\end{table}

\subsubsection{Staff and Barline.}
We symbolize the two staves in piano scores as \textit{R} and \textit{L}. Every score token sequence starts with \textit{R} and is followed by symbols in the right-hand staff; on the other hand, after \textit{L} emerges, symbols in the left-hand staff follow (Figure \ref{fig:token_example}(b)). Whereas, barlines are expressed as \textit{bar} tokens in the same manner as in the MIDI tokenization (Figures \ref{fig:token_example}(b) and (c)).

\subsubsection{Clef and Signatures.}
Clefs, key signatures, and time signatures are tokenized as \textit{clef, key}, and \textit{time} tokens, respectively. Attributes follow each. For example, \textit{clef\_treble} indicates G clef, and \textit{key\_sharp\_2} indicates a key with two sharps, D major or B minor. Besides the beginning of a score, these symbols also appear when the attribute changes (e.g., key change). 

\subsubsection{Voice.}
\label{sec:voice}
A voice, or a stream of notes, is expressed as a pair of tag-like tokens: \textit{<voice>} and \textit{</voice>}. The \textit{<voice>} token declares the start of a new stream, which is followed by the notes and other sym-bols assigned to the voice, whereas \textit{</voice>} indicates the end of the stream. For example, if a measure has two concurrent voices sharing a staff, two sets of voice tokens appear in succession (Figure \ref{fig:token_example}(b)). We represent voices as tag-like tokens to visually clarify the concurrent voices.

\subsubsection{Note and Rest.}
\begin{figure}[h]
 \centerline{
 \includegraphics[width=1.0\columnwidth]{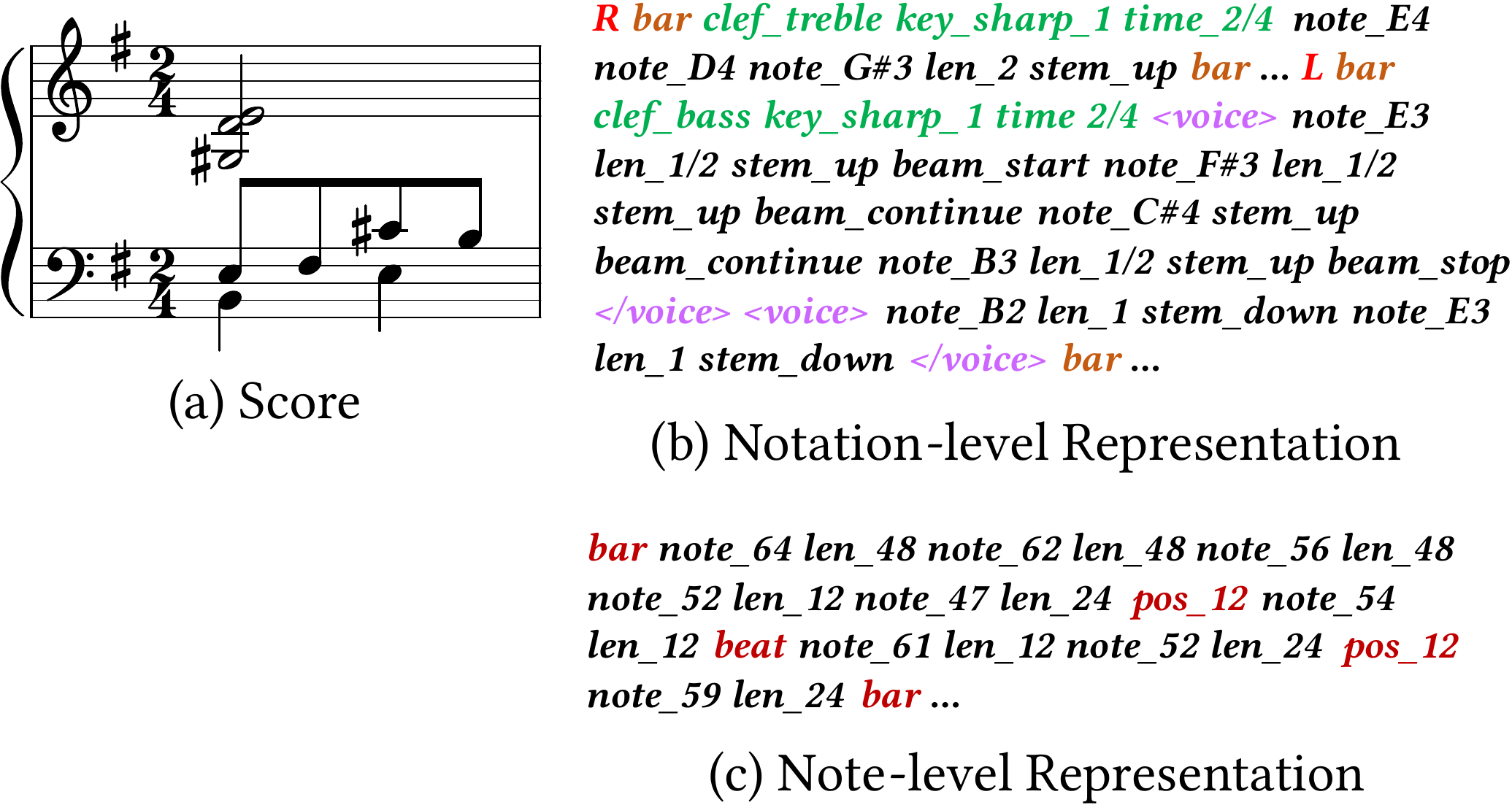}}
 \caption{Example of our token representations corresponding to (a) score excerpt. }
 \label{fig:token_example}
\end{figure}

The note representation consists of five types of tokens that appear in the following order. The \textit{note} token indicates the pitch (e.g., \textit{note\_C4}); the \textit{len} token indicates note duration measured in quarter notes (e.g., \textit{len\_1/2} = 8th note); the \textit{stem} token indicates stem direction (e.g., \textit{stem\_up}); the \textit{beam} token indicates the properties of each beam (e.g., \textit{beam\_start\_continue}, which indicates the initiation of the first beam and the continuation of the second); and the \textit{tie} token indicates the property of tie (e.g., \textit{tie\_stop}). Of these types, \textit{stem, beam,} and \textit{tie} tokens appear only when the preceding note has an attribute. The \textit{note} tokens can appear successively before the \textit{len} token, indicating that the notes constitute a chord (cf. second line in Figure \ref{fig:token_example}(b)).

The rest representation is simple and always consists of two tokens: \textit{rest} and \textit{len}. The \textit{rest} token has no attributes; the subsequent \textit{len} tokens have the same attribute as that of the note duration token (e.g., \textit{len\_1/4}).

\subsubsection{Concatenated Tokens.}
\label{sec:concat}
We consider shortening the score token representation by concatenating tokens because the score token sequences can be relatively lengthy and can thus deteriorate the computational efficiency of training and inference. We concatenate some of the note-attribute tokens because they share large parts of the sequences. By concatenating three note-attribute (duration, stem direction, and beaming) tokens, for example, \textit{len\_1/2}, \textit{stem\_up}, and \textit{beam\_start} are represented as \textit{attr\_1/2\_up\_start} in a concatenated form, where three note-attributes are expressed as a single token.

\subsection{Alternative Approach with Existing Formats}
\label{sec:alternative}
Various formats have been developed for symbolic score representations, such as tree-structured (e.g. MusicXML \cite{Good2003} and MEI \cite{Hankinson2011}) and text-like (e.g. ABC notation, Humdrum (**kern) \cite{Huron2002}, and LilyPond \cite{Nienhuys2003}). Although not designed to feed sequence models, text-like formats can be handled as token sequences when properly segmented. We create token sequences from text-like formats by eliminating header and footer information, and segmenting formatted contents while inserting necessary separators, connectors and staff-voice relation tokens to help restore the formats. For each format, we tokenize note pitch (or rest), note value, and other attributes (e.g., stem direction and beaming group) separately. Meanwhile, for Humdrum, we tokenize its 2D representation in two ways: on a row-major basis and a spine-major basis, denoted as \textit{row} and \textit{spine}, respectively.

\section{From MIDI to Score}
\label{sec:4}

\subsection{Token Representation}
\begin{description}[leftmargin=0em]
\item[Input (MIDI).] For \textit{note}-level tokenization, we adopt REMI \cite{Huang2020}, expanding it by adding \textit{beat} tokens to deal with various meters. With this modification, the position (\textit{pos}) token indicates the note position based on the timing difference from the preceding bar or beat token. We only adopt \textit{bar, beat, position} (\textit{pos}), \textit{note pitch} (\textit{note}), and \textit{note duration} (\textit{len}) tokens for our purpose. A \textit{beat} is evenly divided into 24 timesteps to represent 32nd triplet notes and rests in our setup. An example of note-level tokenization is shown in Figure \ref{fig:token_example}(c).
\\
\item[Output (Score).] We adopt the \textit{notation}-level tokenization methods described in Section \ref{sec:3}. 
\end{description}

\subsection{Model}
In the research area of music generation, many studies have reported that sequence models, such as the Transformer \cite{Vaswani2017}, fit the task quite well \cite{Hsiao2021, Huang2019, Huang2020}. Although many Transformer variants have been proposed \cite{Tay2020}, we adopt the vanilla Transformer model to utilize its original attention mechanism.

\subsection{Training Scheme}
\begin{figure}[h]
 \centerline{
 \includegraphics[width=1.0\columnwidth]{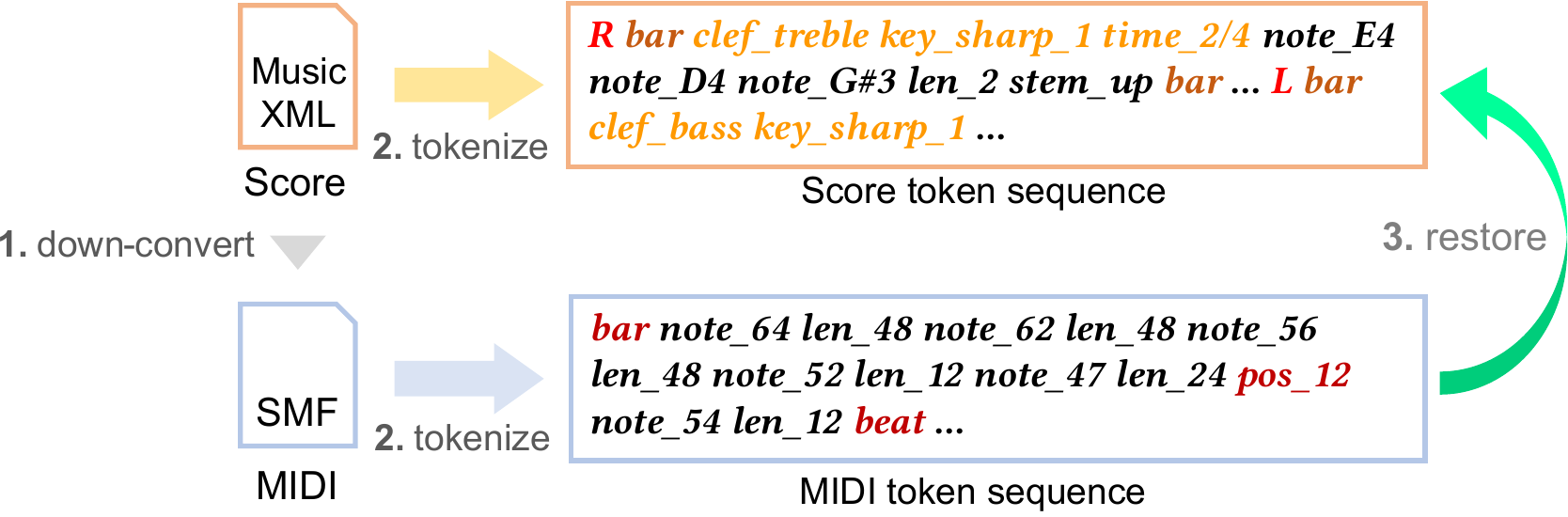}}
 \caption{Training scheme as a score-restoration task. 1) Scores are down-converted into MIDI. 2) Score and MIDI data are tokenized. 3) The model is trained to restore the score from MIDI by estimating lost information.}
 \label{fig:training_scheme}
\end{figure}

Figure \ref{fig:training_scheme} illustrates our training scheme. The model is trained in a supervised manner to restore the original score data from down-converted MIDI data, thus minimizing the loss between estimated and original score tokens. This scheme enables training of the model using only musical score data.

\section{Experimental Setup}
\subsection{Dataset}
We employed popular and classical piano scores for the experiments. For the popular scores, we collected sheet music of popular songs arranged for solo piano from a public website.\footnote{\url{https://www.print-gakufu.com/}}  For the classical scores, we collected piano sheets from KernScores.\footnote{\url{http://kern.ccarh.org/}} The classical data consist of piano sonatas composed by Beethoven, Clementi, Haydn, and Scarlatti, and mazurkas and preludes by Chopin. Table \ref{table:dataset} lists the specifications for each dataset. The scores were split system-by-system (using four measures each for the classical scores) and tokenized into token sequences. We split each dataset song-wise in the ratio 8:1:1 for training, validation, and test data, respectively. Considering the dataset size, we used the popular piano dataset mainly for the experiment and the classical dataset for the cross-genre evaluation.

\begin{table}[h]
 \caption{Specifications of datasets.}
 \label{table:dataset}
 \begin{tabular}{crrr} \toprule
 Genre & \# Pieces & \# Systems & \# Measures \\ \midrule
 Popular piano & 7,161 & 105,132 & 364,129 \\
 Classical piano & 354 & 10,124 & 40,496 \\ \bottomrule
 \end{tabular}
\end{table}

\subsection{Training}
We trained the Transformer model using the following configurations: embedding sizes $d_\mathrm{model} = 256$ and $d_\mathrm{ff} = 512$; number of heads $h = 4$; number of layers of both the encoder and decoder $L = 3$; and dropout rate 0.2. The resulting number of parameters was approximately 4M.

\subsection{Baselines and Alternative Tokenization}
We adopted the transcription framework proposed in \cite{Cogliati2016} as a baseline because it considers musical aspects the most comprehensively to our knowledge among the prior studies. We denote this framework as “CTD.” We used the authors’ implementation with a minor bug fix on a key issue. We also employed the latest version of the music notation software, Finale 26 and MuseScore 3, as baselines. MuseScore 3 has an algorithmic MIDI-to-score functionality,\footnote{\url{https://github.com/musescore/MuseScore}} and Finale 26 has the same functionality. We loaded down-converted single-track MIDI data (with metrical structure) into each software and saved them as MusicXML data.

We also compared the proposed tokenization method with the existing text-like score formats presented in Section \ref{sec:alternative}. For each format, we 1) converted MusicXML scores into the desired format using a converter tool,\footnote{xml2abc, musicxml2hum, and musicxml2ly.}  2) tokenized the output by separating the symbols (see Section \ref{sec:alternative} for details), 3) trained the Transformer model using the token, 4) restored the text in the format by concatenating inferred tokens, and 5) finally reconverted it to MusicXML using another converter tool.\footnote{abc2xml and hum2xml. For LilyPond, we had to scan the LilyPond score because no converter tool worked properly.}

\subsection{Evaluation Metric}
Although most metrics for transcribed musical score quality measure limited musical aspects (e.g., MV2H \cite{McLeod2018}), a metric was proposed in \cite{Cogliati2017} that measures music notation quality based on the number of errors (vs. ground truth score) on 12 musical aspects of a score, utilizing the music21 toolkit \cite{Cuthbert2010}. That study also reported normalized numbers of errors between the transcribed and original scores correlated with subjective evaluation scores. We employed the metric with three expansions: 1) adding three musical aspects (\textit{voice, beam,} and \textit{tie}) to evaluate our model thoroughly and excluding two aspects that were also measured using other aspects (\textit{barline} vs. \textit{time signature}, and \textit{note grouping} vs. \textit{voice}), 2) separating \textit{insertion} and \textit{deletion} errors to discriminate the types of error while integrating \textit{note} and \textit{rest} metrics, 3) counting affected notes and rests for all aspects to facilitate aspect-wise performance comparison (for example, if a \textit{key signature} is wrong, we count notes and rests under its effect). We calculated note-wise error rates based on the metric output, which is the number of errors between the original and generated scores with respect to the 12 musical aspects.

\begin{table*}[ht]
 \caption{Overall error rates in \% (measured based on the difference between the original and generated scores) for the popular piano dataset. Values in boldface show the lowest error rates. \textit{“w/ABC”} uses ABC notation instead of our score token representation and the same for subsequent methods (see Section \ref{sec:alternative}). For \textit{Stem Direction}, models that do not specify directions are excluded from evaluation because the metric cannot measure them properly.}
 \label{table:overall}
 \includegraphics[width=2.1\columnwidth]{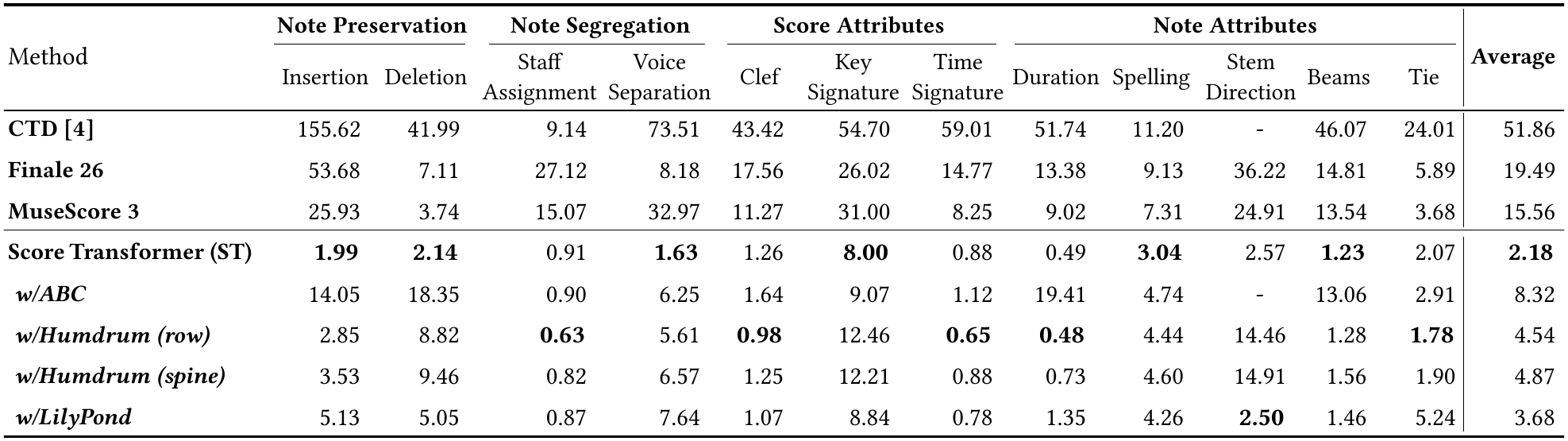}
\end{table*}

\section{Results and Analyses}
\label{sec:6}

\subsection{Comparison with Baselines}
\begin{figure}[h]
 \centerline{
 \includegraphics[width=1.0\columnwidth]{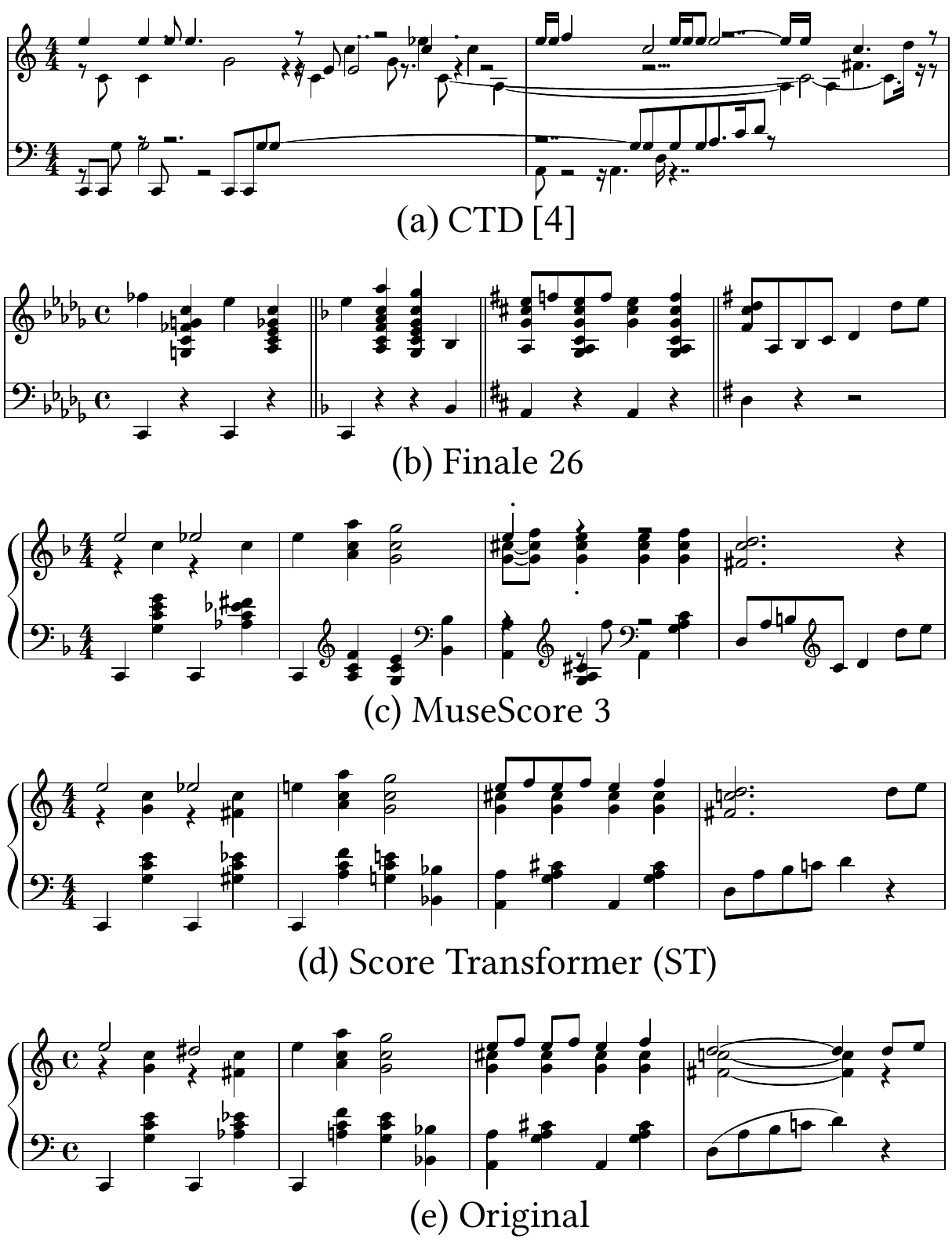}}
 \caption{Generated musical score vs. original score for excerpt of a song that is not in our datasets.}
 \label{fig:output}
\end{figure}

Table \ref{table:overall} shows the overall results for the popular piano dataset measured in error rates for the 12 musical aspects. According to these results, the Score Transformer (ST) performed significantly better than the baseline methods (i.e., CTD \cite{Cogliati2016}, Finale, and MuseScore) on all 12 musical aspects. The results suggest that ST not only successfully learned how to transcribe note-level representations into a musical notation, but also has a much higher capability to notate music than those of prior arts. The results also demonstrate that ST can jointly estimate various symbols and attributes in musical scores while generating non-time-ordered sequences. Figure \ref{fig:output} shows an example of the generated musical scores. Owing to copyright restriction, we used a public-domain song. Although minor differences can be observed, the output of ST was closest to the original score.

\subsection{Evaluation of Alternative Tokenization}
By comparing the error rates of Score Transformer (ST) and its variants with those of existing formats (Section \ref{sec:alternative}) in Table \ref{table:overall} (see lower half), it may be observed that the methods using existing formats exhibit unstable (variably low to high) error rates among the investigated aspects. For example, Humdrum (row) achieved the lowest error rates on five aspects, but it had some of the highest error rates on the other two aspects (i.e., \textit{Key Signature} and \textit{Stem Direction}). The same is true for Humdrum (spine) and LilyPond. In contrast, the proposed tokenization method (ST) demonstrated stable performance over these aspects. The result implies that token representations based on existing formats are not well-adapted for leveraging the attention mechanism of the Transformer.

\begin{figure}[h]
 \centerline{
 \includegraphics[width=1.0\columnwidth]{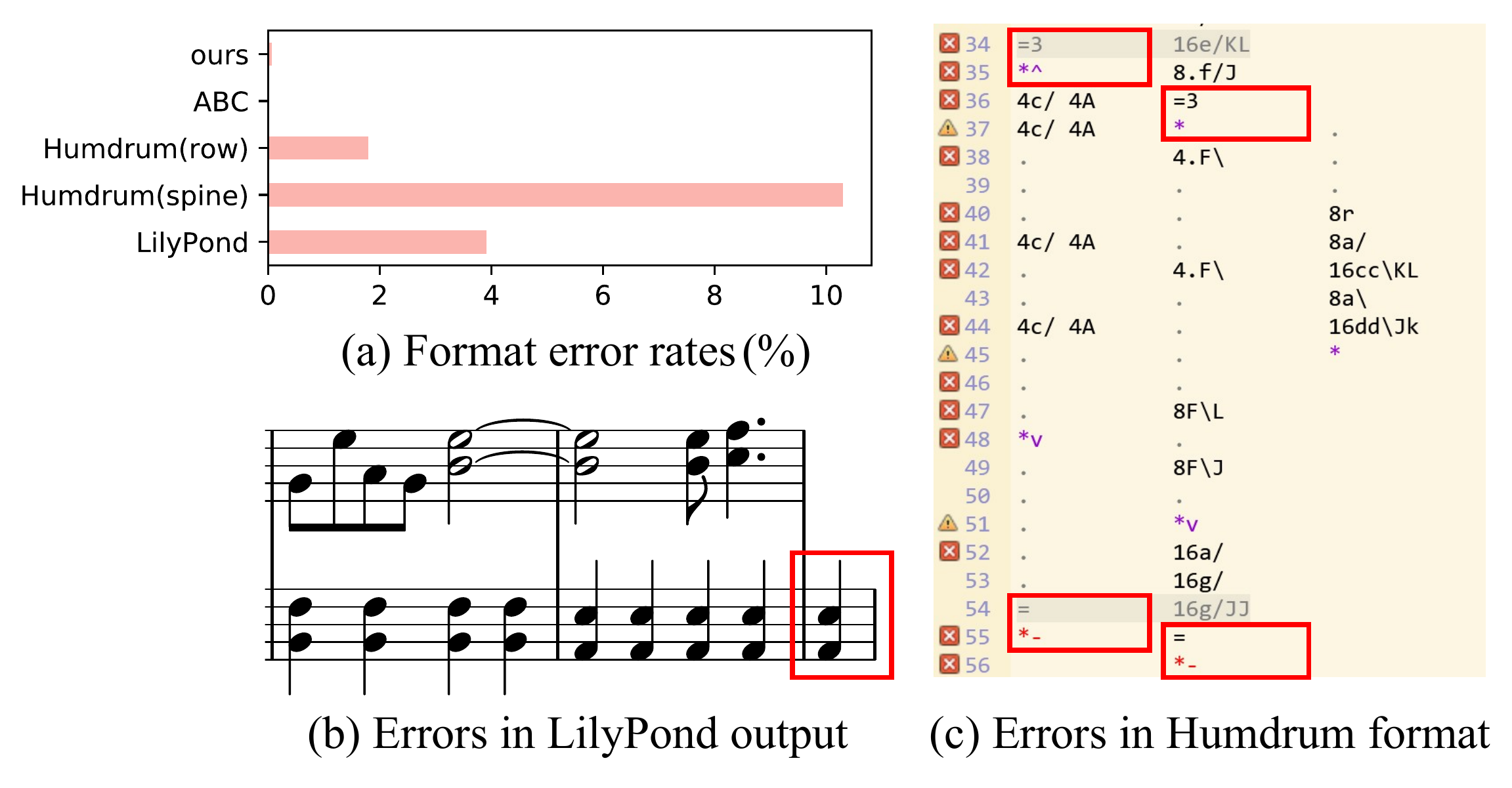}}
 \caption{(a) Rates of format error and frequently observed errors in (b) LilyPond and (c) Humdrum (spine).}
 \label{fig:errors}
\end{figure}

We also determined that some methods are prone to frequent format errors. Figure \ref{fig:errors} shows the rates of format errors and typical errors. Notably, Humdrum (spine) exhibits a high error rate, suggesting that the generation of a 2D Humdrum format is especially difficult for the model when presented on a spine-major (Figure \ref{fig:errors}(c)). For LilyPond, length disagreements between the staves were frequently observed (Figure \ref{fig:errors}(b)). A high error rate in the format leads to practical issues. Based on the measured performance and the format error rates, the proposed tokenization method produced the steadiest results. 

\subsection{Concatenated Form vs. Regular Form}
Table \ref{table:concat} shows the statistics for token length (as average length ± standard deviation) and performance (in error rates) for the regular and concatenated forms (Section \ref{sec:concat}) of our token representation. The average token length indicates the efficiency of the concatenated form in a sequence length (approximately 2/3 of the regular form). The small differences in error rates between the two forms suggest that our score token representation can also be efficient via simple concatenation without resulting in performance degradation.

\begin{table}[h]
 \caption{Comparison between the two forms of score tokens on the popular piano dataset.}
 \label{table:concat}
  \centering
  \footnotesize
  \renewcommand{\arraystretch}{1.15}
 \includegraphics[width=0.95\columnwidth]{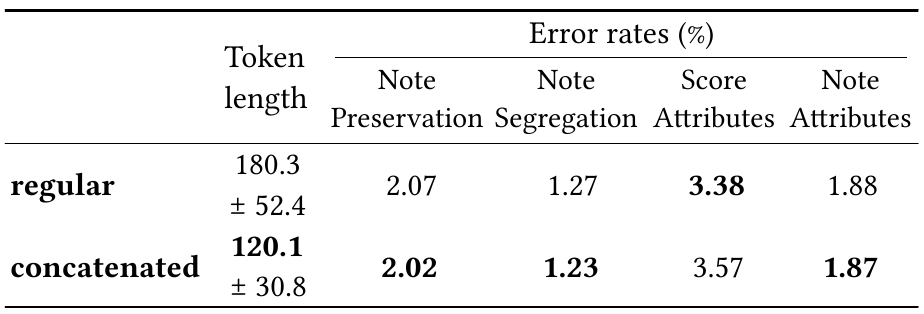}
\end{table}

\subsection{Generalization between Genres}
Table \ref{table:classic} shows the generalization capability of our method. First, the ST model trained on the popular piano dataset (denoted \textit{“pops”}) outperformed baselines on \textit{Average} on the classical piano dataset, showing that it has a certain level of generalization capability. Second, compared with the model trained without the \textit{beat} token in note-level representation (denoted \textit{“pops, w/REMI”}), the \textit{“pops”} model performed better, suggesting that the \textit{beat} tokens in our representation facilitate generalization between genres where the model can encounter the songs with untrained meters.

Table \ref{table:classic} also shows that fine-tuning improves the performance of ST on another genre. From a cross-genre viewpoint, the \textit{“pops”} model performed worse than it did on the popular piano dataset, implying that a performance gap exists between trained and untrained genres. However, following fine-tuning on the classical piano dataset, although ST (\textit{“fine-tuned”}) did not outperform the baselines entirely, the Average error rates of ST decreased by nearly half. The result suggests that ST can perform well also on another genre when fine-tuned on a different genre, even if the dataset is relatively small. 

\begin{table}[h]
 \caption{Error rates (\%) for the classical piano dataset.}
 \label{table:classic}
  \centering
  \footnotesize
  \renewcommand{\arraystretch}{1.15}
    \begin{tabular}{lrrrrr} \toprule
        & \multicolumn{1}{p{4.2em}}{\begin{tabular}{c} Note\\Preservation \end{tabular}} & \multicolumn{1}{p{4.0em}}{\begin{tabular}{c} Note\\Segregation \end{tabular}} & \multicolumn{1}{p{3.5em}}{\begin{tabular}{c} Score\\Attributes \end{tabular}} & \multicolumn{1}{p{3.5em}}{\begin{tabular}{c} Note\\Attributes \end{tabular}} & \multicolumn{1}{p{3.4em}}{\begin{tabular}{c} Average \end{tabular}} \\
    \midrule
    \textbf{CTD [4]} & 104.08  & 55.27  & 67.21  & 40.12  & 55.73  \\
    \textbf{Finale 26} & 27.64  & 32.42  & 41.41  & 24.28  & 29.22  \\
    \textbf{MuseScore 3} & \textbf{10.59} & 30.27  & 23.52  & 19.49  & 19.16  \\
    \midrule
    \textbf{ST \textit{\scriptsize{(pops, w/REMI)}}} & 46.50  & 16.94  & 48.57  & 33.42  & 34.74  \\
    \textbf{ST \textit{\scriptsize{(pops)}}} & 23.12  & 9.79  & 24.89  & 18.87  & 18.44  \\
    \textbf{ST \textit{\scriptsize{(fine-tuned)}}} & 10.87  & \textbf{6.74} & \textbf{14.85} & \textbf{9.08} & \textbf{9.83} \\
    \bottomrule
    \end{tabular}%
    \renewcommand{\arraystretch}{1}
\end{table}

\subsection{Robustness on Unquantized Input}
Finally, Table \ref{table:unquantized} shows the robustness of our method. We experimented with the popular piano dataset with random noise following normal distributions ($\mu=0$, $\sigma=0.08$  for note-on timings; $\mu=0.8, \sigma=0.24$ for note durations) added to all the note-on timings and durations of note-level representation. The sampled values were multiplied by note durations before adding. As a result, the input notes were temporally deviated from the metrical structures. The results for ST were not much worse than those shown in Tables \ref{table:overall} and \ref{table:concat}, suggesting that ST can generate scores robustly, even for noisy input.

\begin{table}[h]
 \caption{Error rates (\%) for unquantized input.}
 \label{table:unquantized}
  \centering
  \footnotesize
  \renewcommand{\arraystretch}{1.15}
    \begin{tabular}{lrrrrr} \toprule
        & \multicolumn{1}{p{4.4em}}{\begin{tabular}{c} Note\\Preservation \end{tabular}} & \multicolumn{1}{p{4.3em}}{\begin{tabular}{c} Note\\Segregation \end{tabular}} & \multicolumn{1}{p{3.7em}}{\begin{tabular}{c} Score\\Attributes \end{tabular}} & \multicolumn{1}{p{3.7em}}{\begin{tabular}{c} Note\\Attributes \end{tabular}} & \multicolumn{1}{p{3.6em}}{\begin{tabular}{c} Average \end{tabular}} \\
    \midrule
    \textbf{CTD [4]} & 130.69  & 41.47  & 59.45  & 52.32  & 65.36  \\
    \textbf{Finale 26} & 29.27  & 17.79  & 23.45  & 23.70  & 23.58  \\
    \textbf{MuseScore 3} & 50.74 & 48.43  & 52.23  & 38.21  & 45.51  \\
    \midrule
    \textbf{ST} & \textbf{3.45} & \textbf{1.82} & \textbf{4.15} & \textbf{2.84} & \textbf{3.10} \\
    \bottomrule
    \end{tabular}%
    \renewcommand{\arraystretch}{1}
\end{table}

\section{Conclusion}
In this paper, we designed score tokens to represent musical scores and trained the Transformer model using paired sequences of \textit{note}-level and \textit{notation}-level data. The evaluation results show that our model significantly outperformed the baseline methods in terms of error rates for a variety of musical aspects. The results clearly show that the Transformer model can generate musical scores with a tokenized representation of scores. Additionally, we demonstrated that our token representation is among the most effective via a comparison with various tokenization methods. Our representation is extendable for various musical symbols and instruments by defining new tokens. We believe that our method opens new possibilities for future research into a variety of musical and multimedia tasks that utilize musical scores.

\begin{acks}
We would like to thank Akira Maezawa and Takuya Fujishima for useful advice and comments on this work.
\end{acks}

\bibliographystyle{ACM-Reference-Format}
\bibliography{ScoreTransformer}

\end{document}